\documentclass[dvips]{article}
\usepackage{icrctc07}

\title{Status of the second phase of the MAGIC telescope}
\shorttitle{MAGIC-II}

\authors{Florian Goebel$^{1}$ \\
For the MAGIC collaboration}
\shortauthors{Florian Goebel et al.}
\afiliations{$^1$Max-Planck-Institut f\"ur Physik, Munich }
\email{fgoebel@mppmu.mpg.de}

\abstract{The MAGIC 17m diameter Cherenkov telescope will be upgraded
  with a second telescope with advanced photon detectors and ultra
  fast readout within the year 2007. The sensitivity of
  MAGIC-II, the two telescope system, will be improved by a factor of 
  2. In addition the energy threshold will be reduced and the
  energy and angular resolution will be improved. 
  The design, status and expected performance of MAGIC-II is presented
  here.}

\begin{document}
\maketitle
%Begin the section.
\section{Introduction}

The 17m diameter MAGIC~\cite{MAGIC} telescope is currently the largest
single dish Imaging Atmospheric Cherenkov telescope (IACT) for very high
energy gamma ray astronomy with the lowest energy threshold among
existing IACTs. It is installed at the Roque de los Muchachos on the
Canary Island La Palma at 2200 m altitude and has been in scientific
operation since summer 2004. Within the year 2007 MAGIC is being
upgraded by the construction of a twin telescope with advanced photon
detectors and readout electronics. MAGIC-II, the two telescope system,
is designed to achieve an improved sensitivity in
stereoscopic/coincidence operation mode and simultaneously lower the
energy threshold. 

All aspects of the wide physics program addressed by the MAGIC
collaboration ranging from astrophysics to fundamental physics will
greatly benefit from an increased sensitivity of the observatory. The
expected lower energy threshold of MAGIC-II will have a strong impact on pulsar
studies and extend the accessible redshift range, which is limited by the
absorption of high energy $\gamma$-rays by the extragalactic
background light. Simultaneous observations with the GLAST satellite,
which will be launched by the end of 2007, will allow detailed studies
of the high energy phenomena in the Universe in the wide energy range
between 100~MeV and 10~TeV.

\begin{figure}
\begin{center}
\includegraphics [width=0.5\textwidth]{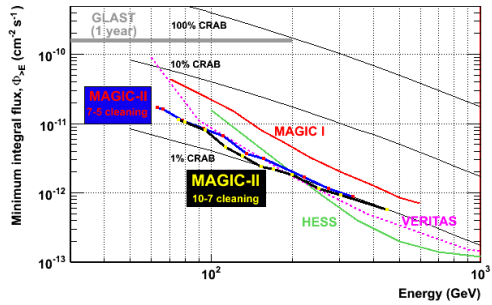}
\end{center}
\caption{MAGIC-II system sensitivity compared with other existing
  experiments (MAGIC-I, HESS and VERITAS). The Crab flux is also shown
  for comparison.}
\label{fig:sensitivity}
\end{figure}

Detailed Monte Carlo studies have been performed to study the expected
performance of MAGIC-II~\cite{MAGICII_MC}. In stereo observation mode,
i.e. simultaneously observing air showers with both telescopes, the
shower reconstruction and background rejection power are significantly
improved. This results in an better angular and energy resolution
and a reduced analysis energy threshold. The overall sensitivity is
expected to increase by a factor of 2 (see figure~\ref{fig:sensitivity}).
Following the results of a dedicated MC study showing moderate
dependence of the sensitivity on the distance of the two telescopes the
second MAGIC telescope has been installed at a distance of 85~m from
the first telescope.

In order to minimize the time and the resources required for design
and production the second MAGIC telescope is in most fundamental
parameters a clone of the first telescope. The lightweight
carbon-fiber epoxy telescope frame, the drive system and the active
mirror control (AMC) are only marginally improved copies of the first
telescope. Both telescopes will be able to reposition within
30-60 min to any sky position for fast reaction to GRB alerts.

\begin{figure}
\begin{center}
\includegraphics [width=0.9\columnwidth,angle=270]{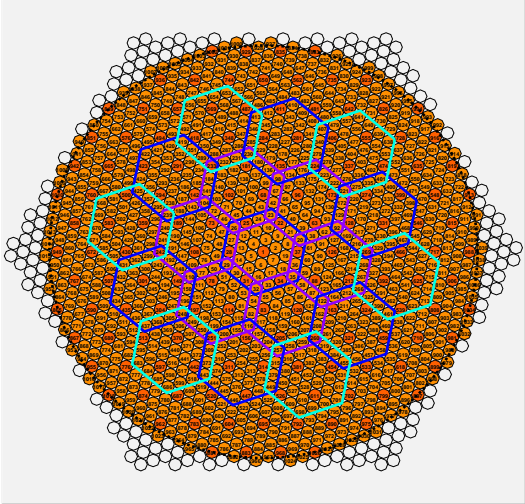}
\caption{Schematics of the MAGIC-II camera. Only colored pixels in a
  round configuration will be equipped. The hexagonal shapes indicate
  the trigger region.}
\end{center}
\label{fig:camera}
\end{figure}

Newly developed components are employed whenever they allow cost
reduction, improve reliability or most importantly increased physics
potential of the new telescope with reasonable efforts. Larger 1~m$^2$
mirror elements have been developed for MAGIC-II reducing cost and
installation efforts. The newly developed MAGIC-II readout system
features ultra fast sampling rates and low power consumption. 
In the first phase the camera will be equipped with increased quantum
efficiency (QE) photomultiplier tubes (PMTs), while a modular camera
design allows upgrades with high QE hybrid photo detectors (HPDs). A
uniform camera with 1039 identical 0.1$^o$ field of view (FoV) pixels
(see figure \ref{fig:camera}) allows an increased trigger area compared
to MAGIC-I.

\begin{figure}
\begin{center}
\includegraphics [width=0.5\textwidth]{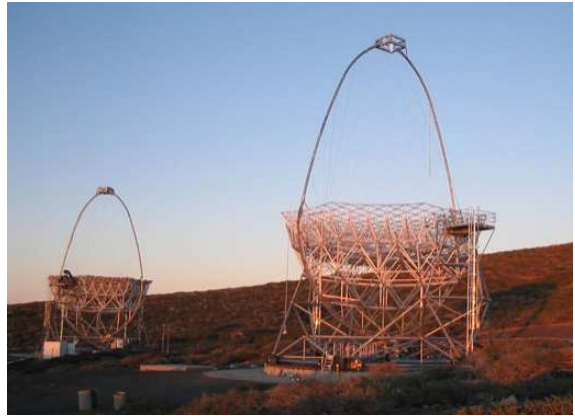}
\end{center}
\caption{In the foreground the second MAGIC telescope. The
  structure is already fully installed. In the background the first
  MAGIC telescope which has been in scientific operation since summer 2004.}
\label{fig:magic2}
\end{figure}

The entire signal chain from the PMTs to the FADCs is designed to have
a total bandwidth as high as 500 MHz. The Cherenkov pulses from
$\gamma$-ray showers are very short (1-2 ns). The parabolic shape of
the reflector of the MAGIC telescope preserves the time structure of
the light pulses. A fast signal chain therefore allows one to minimize
the integration time and thus to reduce the influence of the
background from the light of the night sky (LONS). In addition a
precise measurement of the time structure of the $\gamma$-ray signal
can help to reduce the background due to hadronic background
events\cite{MuxPerformance}.

The fully installed structure of the second MAGIC telescope can be
seen in figure~\ref{fig:magic2}.
In the following the main new developments are discussed.

\section{Mirrors}

Like in MAGIC-I the parabolic tessellated reflector consists of 249
individually movable 1~m$^2$ mirror units, which are adjusted by the
AMC depending on the orientation of the telescope. While in MAGIC-I
each mirror unit consists of 4 individual spherical mirror tiles mounted on
a panel, MAGIC-II will be equipped with 1~m$^2$ spherical mirrors
consisting of one piece.

Two different technologies will be used for the production of the
1~m$^2$ mirrors~\cite{MAGICII_mirrors}. Half of the mirror tiles will
be all aluminum mirrors consisting of a sandwich of two 3~mm thick Al
plates and a 65~mm thick Al honeycomb layer in the center. During
production the sandwich is already bent into a spherical shape,
roughly with the final radius of curvature. The polishing of the
mirror surface by diamond milling is done by the LT Ultra
company. Finally, a protecting quartz coating is applied. The
reflectivity $refl$ and the radius $R_{90}$ of the circle containing
90\% of the spot light have been measured to be around $refl$ = 87\%
and $R_{90}$ = 3~mm.

The other half of the mirrors will be produced as a 26~mm tick
sandwich of 2~mm glass plates around a Al honeycomb layer using a cold
slumping technique. The frontal glass surface is coated with a
reflecting Al layer and a protecting quartz coating. The glass-Al
mirrors show a similar performance as the all Al mirrors.

\section{Camera}

A modular design has been chosen for the camera of the MAGIC-II
telescope~\cite{MAGICII_camera}. Seven pixels in a hexagonal
configuration are grouped to form one cluster, which can easily be
removed and replaced. This allows easy exchange of faulty
clusters. More importantly, it allows full or partial upgrade with
improved photo detectors. The 3.5$^o$ diameter FoV will be similar to
that of the MAGIC-I camera. The MAGIC-II camera will be uniformly
equipped with 1039 identical 0.1$^o$ FoV pixels in a round
configuration (see figure~\ref{fig:camera}).

\begin{figure}
\begin{center}
\includegraphics [width=\columnwidth,angle=0]{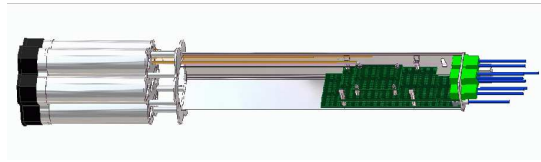}
\end{center}
\caption{Technical drawing of a MAGIC-II camera cluster with 7 pixels.}
\label{fig:cluster}
\end{figure}

In the first phase increased QE PMTs will be used. The Hamamatsu
R10408 6 stage PMTs with hemispherical photocathode typically reach a
peak QE of 34\% \cite{MAGICII_PMTs}. The PMTs have been tested for low
afterpulsing rates (typically 0.4\%), fast signal response ($\sim$1 ns
FWHM) and acceptable aging properties.

%\begin{figure}
%\begin{center}
%\includegraphics [width=\columnwidth,angle=0]{pixel.eps}
%\end{center}
%\caption{Technical drawing of a compact pixel module.}
%\label{fig:pixel}
%\end{figure}

Hamamatsu delivers PMT modules which include a socket with a
Cockcroft-Walton type HV generator. The PMT socket and all the
front-end analog electronics is assembled to form a compact pixel
module. The broadband opto-electronic front-end electronics amplifies
the PMT signal and converts it into an optical pulse, which is
transmitted over optical fibers to the counting house. 

A cluster consists of 7 pixel modules and a cluster body which
includes common control electronics, power distribution and a
test-pulse generator (see figure~~\ref{fig:cluster}). On the front
side the PMTs are equipped with Winston cone type light guides to
minimize the dead area between the PMTs. The slow control electronics
sets the pixel HV and reads the anode currents, the HV values and the
temperature of each pixels. It is in turn controlled by a PC in the
counting house over a custom made RS485 and VME optical link.

Special care has been taken to minimize the weight and the power
consumption of the camera. A water cooling system ensures very good
temperature stabilization.

In a second phase it is planned to replace the inner camera region
with HPDs~\cite{MAGICII_HPDs}. These advanced photo detectors feature
peak QE values of 50\% and will thus significantly increase the
sensitivity for low energy showers. The flexible cluster design allows
field tests of this new technology within the MAGIC-II camera without
major interference with the rest of the camera. Upon successful test
the whole central region of the camera will be equipped with HPDs.

\newpage

\section{Readout}

The optical signals from the camera are converted back to electrical
signals inside the counting house. The electrical signals are split in
two branches. One branch is further amplified and transmitted to the
digitizers while the other branch goes to a discriminator with a
software adjustable threshold. The generated digital signal has a
software controllable width and is sent to the trigger system of the
second telescope. Scalers measure the trigger rates of the individual
pixels. 
 
\begin{figure}
\begin{center}
\includegraphics [width=\columnwidth,angle=0]{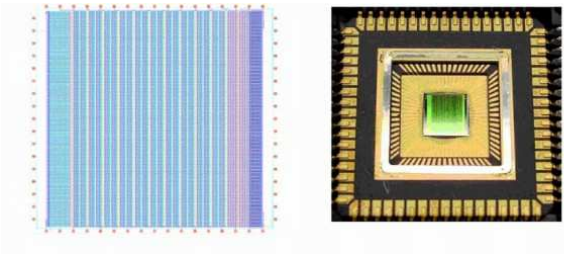}
\caption{Microphotograph of the Domino chip (left) and inside the package (right).}
\label{fig:domino}
\end{center}
\end{figure}

The new 2~GSamples/s digitization and acquisition system is based upon
a low power analog sampler called Domino Ring Sampler (see
figure:~\ref{fig:domino}). The analog signals are stored in a multi
capacitor bank (1024 cell in DRS2) that is organized as a ring buffer,
in which the single capacitors are sequentially enabled by a shift
register driven by an internally generated 2 GHz clock locked by a PLL
to a common synchronization signal. Once an external trigger has been
received, the sampled signals in the ring buffer are read out at a
lower frequency of 40~MHz and digitized with a 12 bits resolution
ADC. The analog sampler, originally designed for the MEG experiment,
has been successfully tested on site and showed a very good linearity
and single photon discrimination capability.

Data management is performed by 9U VME digital boards which handle the
data compression and reformatting as well. Every board hosts 80 analog
channels plus auxiliary digital signals for trigger and monitor
purposes. For a 1~kHz trigger rate and a 2 GHz frequency sampling, the
data throughput can be as high as 100 MBytes/s thus being a challenge for
modern data transmission and storage solutions. The data are
transferred to PCI memory via Gbit optical links using the CERN S-link
protocol and to the mass storage system.

\section{Trigger}
 
The trigger system of the second telescope like the trigger of the
first telescope is based on a compact next neighbor logic. 
However, the uniform camera design allows an increased trigger
area of 2.5$^o$ diameter FoV. This increases the potential to study
extended sources and to perform sky scans. 

When the two telescopes are operated in stereo mode a coincidence
trigger between the two telescopes will reject events 
which only triggered one telescope. This reduces the overall trigger
rate to a rate which is manageable by the data acquisition system.

\section{Acknowledgments}

We would like to thank the IAC for excellent working conditions. The
support of the German BMBF and MPG, the Italian INFN and the Spanish
CICYT, the Swiss ETH and the Polish MNiI is gratefully acknowledged.

%This is the reference to .bib file (Without .bib!)
\bibliography{examplelibrary}
%This in the bibtex style, is ok.
\bibliographystyle{plain}
\end{document}